\documentclass[twocolumn,showpacs,preprintnumbers,amsmath,amssymb]{revtex4}
\usepackage{graphicx}
\usepackage{dcolumn}
\usepackage{bm}
\begin{document}
\title{Resonance State Wave Functions of $^{15}$Be using Supersymmetric Quantum Mechanics}
\author{S. K. Dutta$^1$}
\author{D. Gupta$^2$}
\email{dhruba@jcbose.ac.in}
\author{Swapan K. Saha$^2$}
\affiliation{$^1$Department of Physics, B.G. College, Berhampore, Murshidabad 742101, India}
\affiliation{$^2$Department of Physics, Bose Institute, 93/1 A.P.C. Road, Kolkata 700009, India}
\date{\today}
\begin{abstract}
The theoretical procedure of supersymmetric quantum mechanics is adopted to generate the
resonance state wave functions of the unbound nucleus $^{15}$Be. In this framework, we used 
a density dependent M3Y microscopic potential and arrived at the energy and width 
of the 1.8 MeV (5/2$^+$) resonance state. We did not find any other nearby resonances
for $^{15}$Be. It becomes apparent that the present framework is a powerful tool to theoretically
complement the increasingly important accelerator based experiments with unbound nuclei.
\end{abstract}
\maketitle

\section{Introduction}

\noindent The study of neutron-unbound nuclei are extremely important to probe the neutron-drip 
line in connection with astrophysical problems and other important issues~\cite{BO10,BA12}.
With the increasing availability of exotic rare isotope beams at state of the art
accelerators, researchers are encouraged to carry out studies 
on structure of such unbound nuclei~\cite{SN13}. To complement such 
experimental findings, robust theoretical frameworks are indispensable.\\ 

\noindent The beryllium isotopic chain is a very attractive arena
to make a systematic study extending from stable nuclei to its neutron-unbound rare 
isotopes~\cite{KU15}. The first attempt to populate the unbound nucleus 
$^{15}$Be was carried out in 2011
by Spyrou et al~\cite{SP11} using a two proton knockout reaction from a $^{17}$C 
secondary beam. The search for a resonance was unsuccessful and it was concluded
that $^{15}$Be is unbound by more than 1.54 MeV. Attempts were made to predict states of
$^{15}$Be in the shell-model framework~\cite{WA92,BR07,PO85}. The shell-model 
calculations in Ref.~\cite{WA92,BR07} predicted a $\frac{3}{2}^{+}$ ground state for $^{15}$Be
together with a low lying $\frac{5}{2}^{+}$ excited state at approximately 300 keV and
additional states above 1.2 MeV whereas Ref.~\cite{PO85} predicted just the opposite, 
namely a $\frac{5}{2}^{+}$ ground state with an excited $\frac{3}{2}^{+}$ state at 70 keV. 
Clearly, the existing theoretical calculations of $^{15}$Be level structure, in the 
shell-model framework, is very uncertain. To observe the $^{15}$Be nucleus,
Snyder et al~\cite{SN13} carried out a $^{14}$Be(d,p)$^{15}$Be reaction in 2013 leading
to the first observation of $^{15}$Be. It is pertinent that the selected channel 
is expected to populate
both the $\frac{3}{2}^{+}$ and $\frac{5}{2}^{+}$ states in $^{15}$Be. The $^{15}$Be, produced 
in this neutron transfer reaction, being unbound decayed immediately into $^{14}$Be 
and a neutron. The reconstructed decay energy spectrum exhibits a resonance
at 1.8(1) MeV and the spin parity of this state is tentatively assigned to be $\frac{5}{2}^{+}$.
However, the experiment does not resolve the question of which of the two states among
$\frac{3}{2}^{+}$ and $\frac{5}{2}^{+}$ corresponds to the ground state. More recently 
in 2015, Kuchera et al~\cite{KU15} performed a two proton knockout reaction using 
$^{17}$C secondary beam on thick $^9$Be target to find the predicted $^{15}$Be 
$\frac{3}{2}^{+}$ state. However, the results indicate that a $^{15}$Be component was not 
needed to describe the data. Thus, we see that experiments
on $^{15}$Be are very few and there are substantial ambiguities in the results of 
theoretical calculations relevant to experimental findings.\\

\section{Theory}

\noindent In the present work, we resort to a very effective technique of supersymmetric 
quantum mechanics (SQM) for the study of unbound nuclear systems. Earlier,
we successfully applied SQM to detect low-lying broad resonances of the weakly bound 
nucleus $^{11}$Be~\cite{DU14}. The effectiveness of our
theoretical procedure is due to its ability to circumvent the numerical challenges posed by 
the shallow potential of such nuclear systems. 
The study of resonance states of $^{15}$Be is a challenging problem as it has no bound state
while its two-body potential for a resonance state is a shallow well, followed by a very 
low and wide barrier. The low barrier does temporarily trap the system leading to a
broad resonance. This inadvertently hinders accurate calculation of resonance energy masked 
by the broad resonance width.  Under the circumstances, we decided to adopt
SQM to study the resonance states of the unbound nucleus $^{15}$Be.  We treat it in the 
framework of a two-body model consisting of an inert core of $^{14}$Be and a single valence 
neutron. The density distribution used for 
$^{14}$Be consists of separate core and halo components. The parameterization used 
is named Gaussian-Halo, reflecting an rms radius of 3.1 fm for $^{14}$Be~\cite{IL09}.
The two-body potential $v(r)$ is generated microscopically in a single folding model 
using the density dependent M3Y (DDM3Y) effective interaction~\cite{DU14}. 
The DDM3Y interaction was used also in our earlier work~\cite{DU14}, to study the weakly bound
nucleus $^{11}$Be. In this context, it may be noted that 
the DDM3Y effective interaction was found earlier also to give a satisfactory description of 
radioactivity, nuclear matter and scattering~\cite{BA04,BA05,GU06}.\\

\noindent From the constructed microscopic potential, SQM generates a family of isospectral 
potentials (IP),
which have a normalizable positive energy solution at a selected energy. This is a lesser known 
result of SQM, namely a bound state in the continuum (BIC)~\cite{DA82,NI84,KH89,PA93}. This IP has desirable 
properties which can be utilized to extract information about unbound resonance states. The 
microscopic potential constructed from single folding calculation is in general a shallow well 
followed by a low and wide barrier. For a finite barrier height, in principle, a system may be 
temporarily trapped inside the shallow well when its energy is close to the resonance energy. 
In reality, there is a very high probability for tunnelling through the barrier
which gives rise to broad resonance widths. Our technique bypasses this problem and obtains 
accurate
resonance energies. This is achieved by construction of isospectral potentials that are deep 
enough to supress the tunnelling probability of the system and yet provide an accurate location 
of resonance states. This becomes even clearer by the plots of the resulting
resonance state wave function.\\

\noindent Going further, an isospectral partner potential could be constructed by following the ideas extended
by Pappademos {\it et al}~\cite{PA93} to scattering states with positive energy in the continuum. 
Wave functions in the continuum are non-normalizable but following~\cite{PA93}, one 
can construct normalizable wave functions at a selected energy, which represents a BIC. The BIC 
represents a solution of the equation with an isospectral potential $\hat{v}(r;\lambda)$, 
where $\lambda$ is a parameter which affects the strength of IP~\cite{DU14}. It follows from theory
as well as in practice that resonance energy does not
depend on the choice of $\lambda$. So a suitable choice of $\lambda$ ensures the stability of the
resonance state. It preserves the spectrum of the original potential while adding a discrete BIC 
at a selected energy.\\ 

\noindent As already mentioned, we have considered quasi-bound $^{15}$Be to be a two-body system ($^{14}$Be + $n$), and for microscopically constructed
$v(r)$, inclusive of the centrifugal barrier,
it is possible to construct a family of strictly isospectral potentials $\hat{v}(r;\lambda)$ for arbitrarily chosen
parameter $\lambda$. This is done by the BIC formalism as follows. The equation for
positive energy resonance state wave function $\psi_{E}(r)$ is
\begin{equation}
\left(-\frac{\hbar^{2}}{2\mu}\frac{d^{2}}{dr^{2}} + v(r) - E
\right) \psi_{E}(r) = 0
\label{isoeq}
\end{equation}
where $E$ is the energy of the resonance state. Here we follow the procedure described in~\cite{PA93} and
solve the two-body Schr\"odinger equation for a positive energy $E$ subject to the
boundary condition $\psi_{E}(0)=0$ and normalized to a constant (fixed) amplitude of oscillation in the
asymptotic region. The solution $\psi_{E}(r)$ is not square integrable and it oscillates as r increases.
It can be verified by direct substitution that
$\hat{\psi}_{E}(r;\lambda)$ 
satisfies Eq.~(\ref{isoeq}), where

$\hat{\psi}_{E}(r;\lambda) = \frac{\psi_{E}(r)}{I_{E}(r) + \lambda} \:$,
and
$ I_{E}(r) = {\displaystyle\int}_{0}^{r}{[\psi_{E}(r^{\prime} )]}^
{2}dr^{\prime}$
with $v(r)$ replaced by

\begin{equation}
\hat{v}(r;\lambda) = v(r) - \frac{\hbar^2}{2\mu}\left[\frac{4\psi_{E}(r)\psi_{E}^
{\prime}(r)}{I_{E}(r) + \lambda}
+\frac{2(\psi_{E}(r))^{4}}{{(I_{E}(r) + \lambda)}^{2}}\right] \:,
\label{vhat}
\end{equation}

\noindent Now the potential $\hat{v}(r;\lambda)$ given by Eq.~(\ref{vhat})
depends on the arbitrary parameter $\lambda$, $\hat{\psi}_{E}(r)$ is
the solution to the energy $E$. As $\psi_{E}(r)$ oscillates with a constant amplitude in the 
asymptotic region, it is observed $I_{E}(r)$ increases approximately linearly with $r$ for
large $r$ and from above equations we find
$\hat{\psi}_{E}(r;\lambda)$ to be normalizable. Thus
$\hat{\psi}_{E}(r;\lambda)$ represents a BIC of $\hat{v}(r;\lambda)$, which is isospectral with
$v(r)$. Hence $\hat{v}(r;\lambda)$ develops a
deep and narrow well followed by a high barrier which advances towards the origin for
$\lambda\rightarrow0+$ and approaches  $v(r)$ for $\lambda\rightarrow+\infty$.\\

\noindent The deep well and high barrier combination effectively traps the system giving rise 
to a quasi-bound state. We calculate the
probability of the system to be trapped within this enlarged well-barrier combination as
\begin{equation}
C(E)={\displaystyle\int}_{r_{a}}^{r_{b}} {[\hat{\psi}_{E}(r^{\prime})]}^{2} dr^{\prime}  \;,
\label{ce}
\end{equation}
where $r_{a}$ and $r_{b}$ are radial distances at the classical turning points $a$, $b$ within the
potential well.
Our method is advantageous when it comes to highly accurate calculation of resonance energy 
along with numerical ease. In our procedure, the probability $C(E)$ of the system to be trapped in well-barrier combination of $\hat{v}(r;\lambda)$ shows a sharp peak at the resonance energy 
for appropriate choice of $\lambda$. Resonance state in the original potential $v(r)$ is 
apparent but not prominent within the well. It gets enhanced by a very large amount in 
$\hat{v}(r;\lambda)$ giving rise to a sharp peak in probability $C(E)$. Although the probability 
plots $C(E)$ of the system exhibits independence of resonance energy on the choice of $\lambda$ 
values, eventually a judicious choice of $\lambda$ is necessary to eliminate numerical errors 
in the wave function.\\

\noindent Width ($\Gamma$) of the resonance is calculated from its mean life
($\tau$), using the time energy uncertainty relation. The mean life is given by the
reciprocal of the decay constant, which is expressed as a product of the number
($n_{c}$) of impacts on the barrier and the transmission probability ($T$).
Semiclassical estimation of $n_{c}$  is obtained as the reciprocal of time of flight
within the well of $\hat{v}$ between the classical turning points $a$ and $b$.
Here, $T$ is given by the WKB approximation for the transmission through
the barrier of $\hat{v}$ with resonance energy $E_{R}$,
\begin{equation}
T= exp[-2 {\displaystyle\int}_{b}^{c}
\sqrt{\frac{2\mu}{\hbar^2}(\hat{v}(\lambda;r)-E_{R}})  dr]
\end{equation}
where $b$ and $c$ are the classical turning points of the barrier. The final
expression for $\Gamma$ is
\begin{equation}
\Gamma=2\:\:\sqrt{\frac{\hbar^2}{2\mu}}\:\:\frac{ exp(-2
{\displaystyle\int}_{b}^{c}
\sqrt{\frac{2\mu}{\hbar^2}(\hat{v}(\lambda;r)-E_{R}})  dr)}
{{\displaystyle\int}_{a}^{b} \frac{dr}{\sqrt{(E_{R} -
\hat{v}(\lambda;r))}}}. \label{gamma}
\end{equation}

\noindent We have verified by direct calculation that $\Gamma$ is
independent of $\lambda$ within numerical errors.\\

\begin{figure}
\includegraphics[scale=0.50]{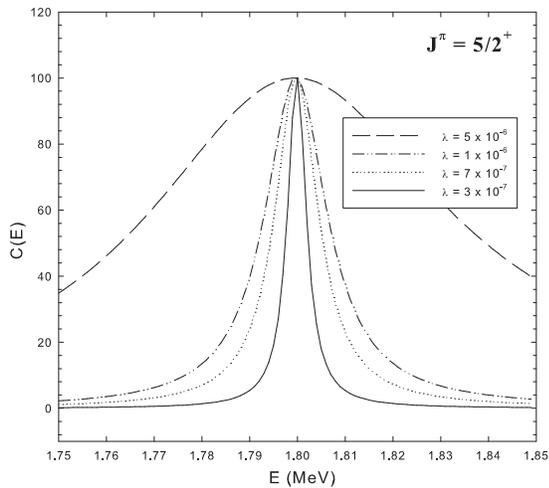}
\caption{\label{fig1}  Probability $C(E)$ as a function of energy $E$ for $\lambda=5 \times 10^{-6},
1 \times 10^{-6}, 7 \times 10^{-7} {\rm ~and~} 3 \times 10^{-7}$ for the $\frac{5}{2}^{+}$ state of  $^{15}$Be.}
\end{figure}

\begin{figure}
\includegraphics[scale=0.50]{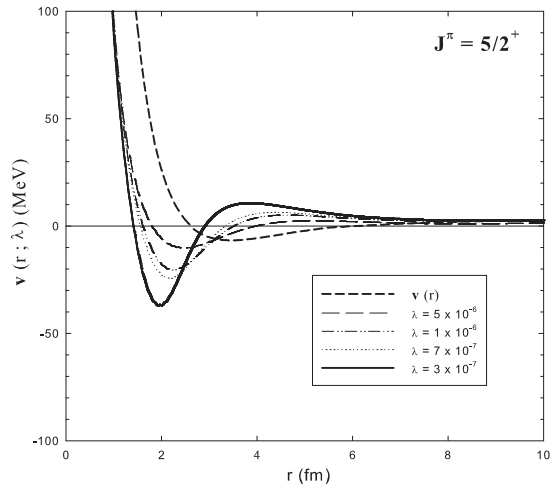}
\caption{\label{fig2} One parameter family of isospectral potentials $V(\lambda;r)$ for 
$\lambda=5 \times 10^{-6},
1 \times 10^{-6}, 7 \times 10^{-7} {\rm ~and~} 3 \times 10^{-7}$ for the $\frac{5}{2}^{+}$ state of  $^{15}$Be.}
\end{figure}

\begin{figure}
\includegraphics[scale=0.50]{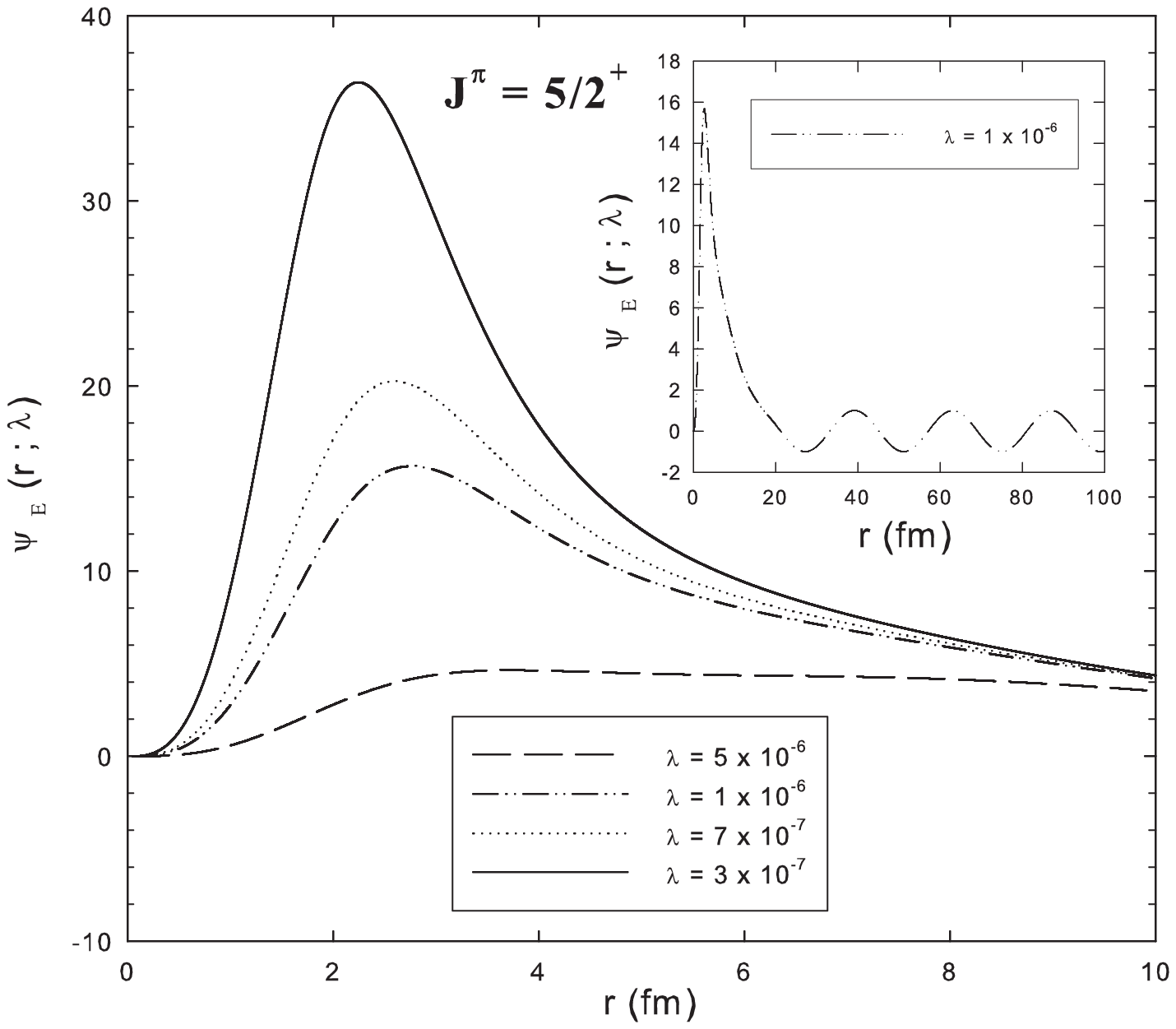}
\caption{\label{fig3} Wave function (in arbitrary units) at the excitation energy of
1.80 MeV for $\lambda = 5 \times 10^{-6},
1 \times 10^{-6}, 7 \times 10^{-7} {\rm ~and~} 3 \times 10^{-7}$ for the $\frac{5}{2}^{+}$ state 
of  $^{15}$Be. The inset shows the wave function plot for $\lambda = 1 \times 10^{-6}$ in an 
expanded scale up to 100 fm.}
\end{figure}

\begin{figure}
\includegraphics[scale=0.50]{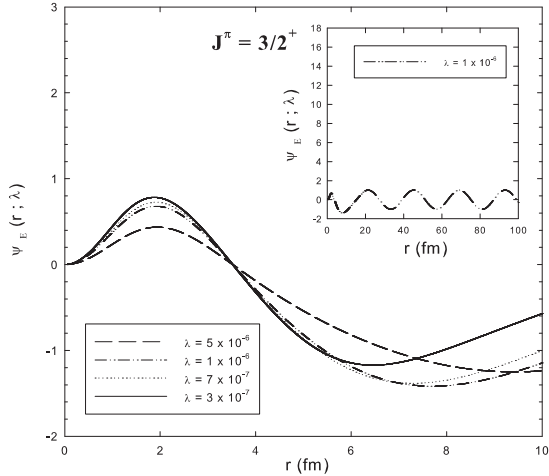}
\caption{\label{fig4} Wave function (in arbitrary units) at the excitation energy of
1.80 MeV for $\lambda = 5 \times 10^{-6},
1 \times 10^{-6}, 7 \times 10^{-7} {\rm ~and~} 3 \times 10^{-7}$ for the $\frac{3}{2}^{+}$ state 
of  $^{15}$Be. The inset shows the wave function plot for $\lambda = 1 \times 10^{-6}$ in an 
expanded scale up to 100 fm.}
\end{figure}

\section{Results}

\noindent A plot of $C(E)$ as a function of $E$ for various $\lambda$ 
values shows how the trapping effect of $\hat{v}(r;\lambda)$ increases as $\lambda$ decreases. 
For appropriate choice of $\lambda$, suitable isospectral potentials are constructed. Probability $C(E)$ of the system for $\frac{5}{2}^{+}$ state is plotted in Fig. 1. We present $C(E)$ for 
$\lambda=5 \times 10^{-6}, 1 \times 10^{-6}, 7 \times 10^{-7} {\rm ~and~} 3 \times 10^{-7}$ in 
the same figure, normalizing each curve to a peak value of 100. Sharpness of the peak increases 
rapidly as $\lambda$ decreases towards smaller values. It is evident from the plot that there is 
a resonant state at energy $E$ = 1.8 MeV for $^{15}$Be. The Fig. 2 shows the family of 
isospectral potentials $\hat{v}(r;\lambda)$ at resonance energy $E_{R}$ for the same  
$\lambda$ values as in Fig. 1 for the $\frac{5}{2}^{+}$ state of $^{15}$Be along with the original 
single folded potential $v(r)$. Enhancement of trapping probability of the system for the 
$\frac{5}{2}^{+}$ state of  $^{15}$Be  as $\lambda$ decreases could be seen from Fig. 3. Wave functions $\hat{\psi}_{E}(r;\lambda)$ at the resonance energy
1.8 MeV for the same $\lambda$ values as in Fig. 1 are presented in Fig. 3. The 
asymptotic region represents a free particle while a peak appears within the well-barrier 
combination, representing the enhanced probability of the particle being trapped inside the well. 
This peak increases rapidly as  $\lambda$ decreases towards lower values. The inset of Fig. 3 shows the wave function plot for $\lambda = 1 \times 10^{-6}$ in an expanded scale up to 100 fm.
These wave function plots are signatures of resonance states. They display appreciable 
amplitude within the well and behave as free particles exhibiting sinusoidal wave function in 
asymptotic region once it leaks out of the well-barrier trapping effect. We tried the same 
procedure to locate the presence of $\frac{3}{2}^{+}$ state. There was no confirmation regarding
any contribution from $\frac{3}{2}^{+}$ state as evident from Fig. 4. Within the well, the low
amplitude of the $\frac{3}{2}^{+}$ wave function in comparison to the $\frac{5}{2}^{+}$ state
and in asymptotic region nearly similar oscillatory behavior of the two states justifies our point.
Wave functions in Fig. 4 have small amplitude in the well and sinusoidal nature 
in asymptotic region, representing an unbound state. 
Plot of wave function for $\frac{3}{2}^{+}$ state of $^{15}$Be clearly rules out any presence of 
resonance state.  A similar search for $\frac{7}{2}^{+}$ 
resonance state was carried out and its wave functions represented unbound system with no resonance effect.\\

\noindent The accuracy in locating the resonance energy $E_{R}$ could be increased by the 
choice of an optimum
value of $\lambda$, although in general $E_{R}$ is independent of $\lambda$. This optimized 
$\lambda$ value could be used in all further calculations cutting down numerical computational 
time. The procedure adopted to study resonances in binary systems can also be used in calculating 
differential cross sections as a function of energy. Wave functions for resonance states are 
readily available in our procedure which could be advantageous for further extended calculations.
The resonance width obtained is $\Gamma = 470$ keV as compared to the 
experimental finding~\cite{SN13} of $\Gamma_{exp}$ = 575 (200) keV. We also found that $\Gamma$ is independent of $\lambda$.\\

\noindent The novelty of our method lies in the extraction of resonant state wave functions 
with the help of SQM procedure. We would like to reiterate that the isospectral potentials 
are generated 
from an effective potential $v(r)$ which represent the same energy quasi-bound state. The wave 
function shown in the inset of Fig. 3 establishes firmly the quasi-bound nature of the state. 
All the wave functions have the same nature namely a high probability inside the potential 
well-barrier and free particle nature once it tunnels out from the trap. The depth and width of
the isospectral potential well, and the height and width of the barrier adjust itself in such
a way that they are able to reproduce the resonance energy as well as the  
width. Since isospectral potentials could represent quasi-bound 
states and reproduce the quasi-bound state energies, it is not surprising to expect that they 
will represent another property of such states viz. width of the states. Our calculation also
points out that different shaped wave functions modified by $\lambda$ not only represent 
the position of resonance but also the width of the state.\\

\section{Conclusion}

\noindent In conclusion, we have been able to generate the wave functions in the SQM 
framework with a DDM3Y microscopic potential and arrive at the unbound state energy and width, 
of the $^{15}$Be nucleus. Our procedure confirmed the existence of $\frac{5}{2}^{+}$
state and its experimentally observed unbound resonance energy~\cite{KU15}.
In the present formalism, though $\lambda$ appears as a parameter to enhance resonance effect
but it has no role in locating the exact resonance energy. Excellent agreement with the 
experimental results for $^{15}$Be can be ascribed to the realistic two-body 
($^{14}$Be + $n$) folded potential. The SQM is
the only procedure by which resonant state wave functions are extracted and utilized to 
effectively reproduce an experimental observable ($\Gamma$). Wave functions of same $J^{\pi}$ 
but different $\lambda$ values are in a sense equivalent as they reproduce the same resonance
energy and width of the state.\\ 

\noindent Physics of exotic unbound nuclei would 
pervade the field of nuclear physics in the coming years and a robust theoretical 
framework exclusively to study such nuclei is necessary. Earlier, we successfully applied
this procedure to an unstable nucleus $^{11}$Be. The present work shows that the same
procedure works also on an unbound nucleus with excellent results.\\

\end{document}